\documentclass[11pt]{article}
\usepackage{moriond,epsfig}

\def\Journal#1#2#3#4{{#1} {\bf #2}, #3 (#4)}

\def\APJ{ ApJ}
\def\APJS{ApJS}
\def\MN{MNRAS}
\def\arxiv{arXiv:astro-ph}

\def\be{\begin{equation}}
\def\ee{\end{equation}}
\def\bea{\begin{eqnarray}}
\def\eea{\end{eqnarray}}

\begin{document}
\vspace*{4cm}
\title{Investigating The Uncertainty On The BAO Scale Measured From Future Photometric And Spectroscopic Surveys}

\author{Alexandra Abate, Reza Ansari \& Marc Moniez}
\address{Laboratoire de l'Acc\'{e}l\'{e}rateur Lin\'{e}aire, IN2P3-CNRS, Universit\'{e} de Paris-Sud, BP. 34, 91898 Orsay Cedex, France}

\author{Alexia Gorecki, Aur\'{e}lien Barrau, Sylvain Baumont \& Laurent Derome}
\address{Laboratoire de Physique Subatomique et de Cosmologie, CNRS/UJF/INPG 53, rue des Martyrs 38026 Grenoble Cedex, France}

\maketitle\abstracts{
The Large Synoptic Survey Telescope (LSST) is a wide (20,000 sq.deg.) and deep $ugrizy$ imaging survey which will be sited at Cerro Pachon in Chile.  A major scientific goal of LSST is to constrain dark energy parameters via the baryon acoustic oscillation (BAO) signal. Crucial to this technique is the measurement of well-understood photometric redshifts, derived from the survey \textit{ugrizy} imaging. Here we present the results of the effect of simulated photometric redshift (PZ) errors on the reconstruction of the BAO signal.   We generate many ``Monte Carlo" simulations of galaxies from a model power spectrum using Fast Fourier Transform techniques.  Mock galaxy properties are assigned using an algorithm that reproduces observed luminosity-color-redshift distributions from the GOODS survey.  
We also compare these results to those expected from a possible future spectroscopic survey such as BigBOSS.}

\section{Introduction}
In the early universe, when the temperature is high enough so the photons and baryons are coupled through Compton scattering in a plasma, the cosmological density fluctuations create sound waves which propagate through this plasma.  At around a redshift of 1000 the temperature of the universe drops to a level at which the Compton scattering is no longer efficient, effectively stalling the sound waves at the epoch of recombination.  The distance these sound waves could travel in the time between the formation of the perturbations and the epoch of recombination imprinted a \textit{characteristic scale} into the spectrum of density perturbations.  Because the universe has a significant fraction of baryons, theory predicts that this characteristic scale will also be imprinted in the late-time spectrum of density perturbations.  Since galaxies are expected to form in the regions that are overdense in baryons and dark matter, and because this is driven by where the initial perturbations were, there should be an small excess of galaxies at this characteristic scale away from other galaxies.  At recombination this scale is roughly 150 Mpc, and it appears in the power spectrum of density fluctuations as a damped harmonic sequence, a series of \textit{wiggles}, which are what is known as the baryon acoustic oscillations (BAO).
The position of these wiggles measured as a function of redshift reveals information about dark energy.


It has been shown\cite{bb05,bmegaz} that it is possible to measure BAO using a photometric redshift survey.  The advantage of using photometric redshifts over spectroscopic redshifts is that they are much less expensive (time consuming) to obtain and so a much larger volume of the universe can be surveyed.  Since the BAO feature is observed on fairly large scales, a volume even larger than this must be surveyed in order to obtain a measurement with enough statistical significance to measure the cosmological parameters precisely.  The disadvantage of photometric redshifts is that they have large errors and the exact distribution of these errors is currently not well understood.  Here we present an early result from our simulations of the performance of LSST, a future photometric wide-field survey, in measuring the BAO signal.

\section{The Large Synoptic Survey Telescope}

The Large Synoptic Survey Telescope (LSST) will be the first of a next generation of ground based optical telescopes. 
It seeks to investigate important scientific problems of the next decade: probing dark energy and dark matter, taking an inventory of the Solar System, exploring the transient optical sky, and understanding galaxy formation and the structure of the Milky Way.  All these diverse science goals require wide-field repeated deep imaging of the sky in many optical bands.

LSST will carry out such a survey by imaging 20,000 deg$^2$ of the sky in six broad photometric bands $ugrizy$, imaging each region of sky roughly 2000 times (1000 pairs of back-to-back 15-sec exposures) over a ten-year survey lifetime.  After 10 years, with 1000 visits (2x15s exposures), it is anticipated LSST will yield a coadded map of the sky with a 5-$\sigma$ depth for point sources of $r\sim 27.5$.  For more details on the science goals and capabilities that will be provided by LSST, see the LSST Science Book\cite{lsstsb}.

\section{Simulation of mock galaxy catalogs}
\label{sec:sim1}
\begin{figure}
\center
\psfig{figure=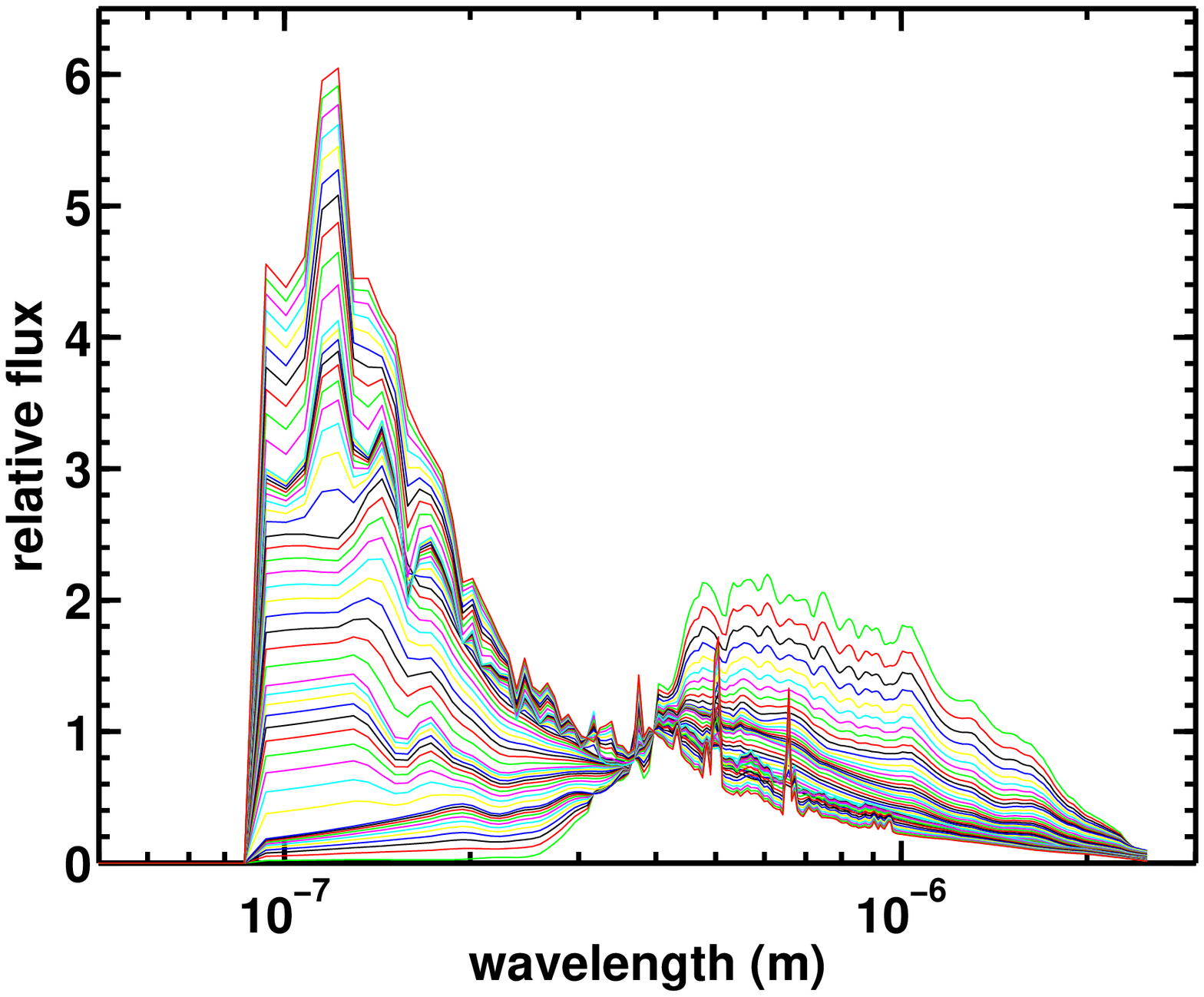,height=1.5in}
\psfig{figure=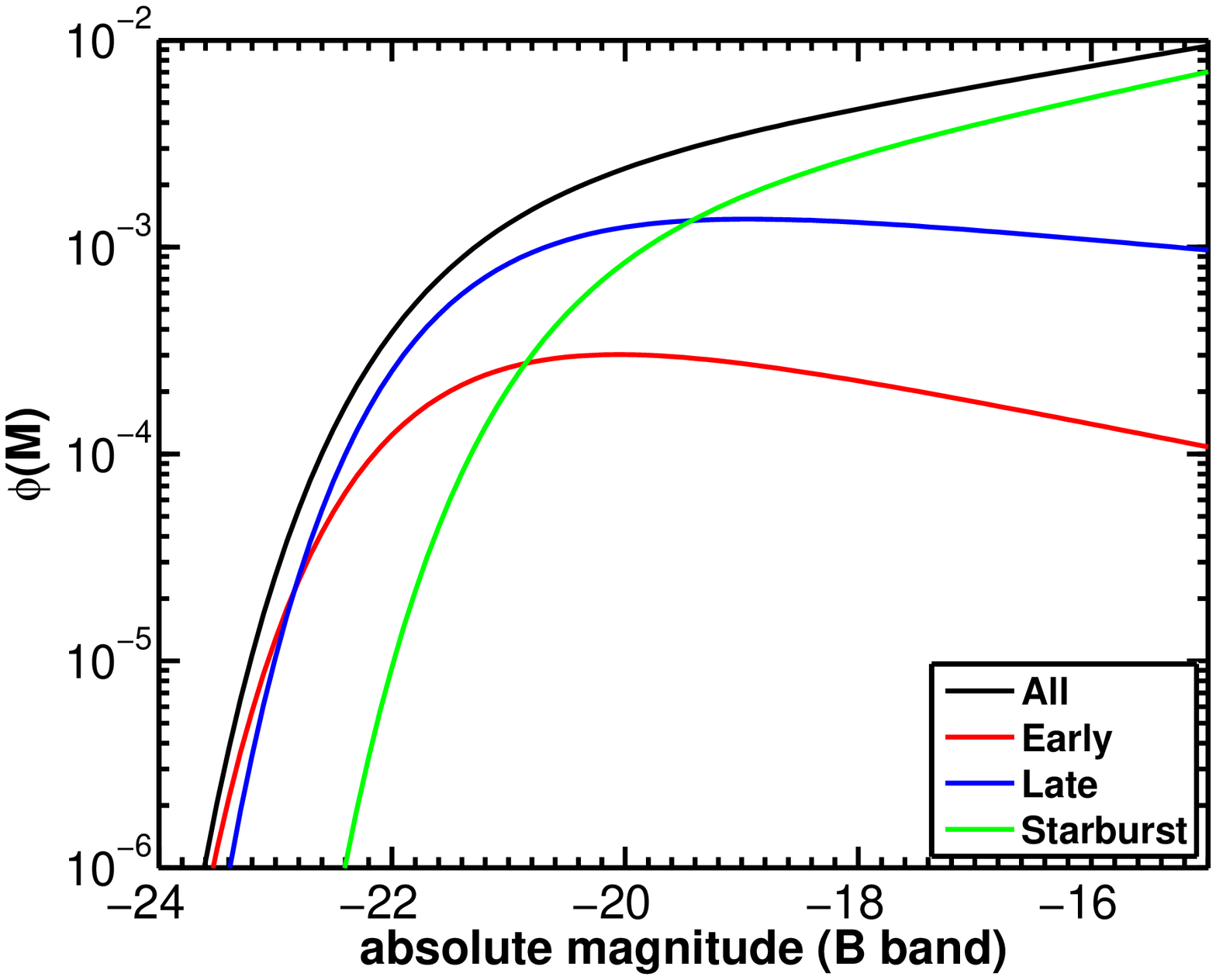,height=1.5in}
\psfig{figure=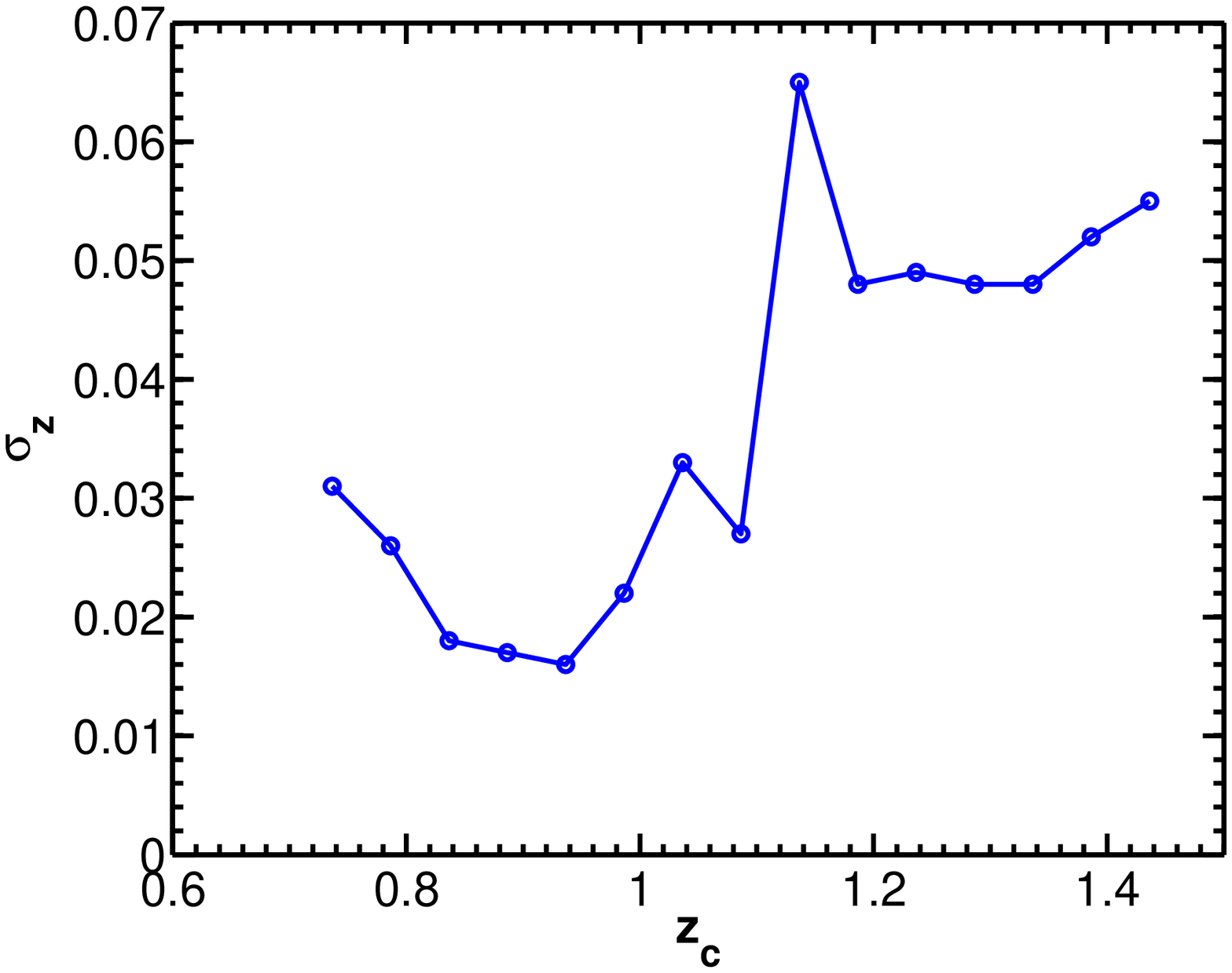,height=1.5in}
\caption{\textit{LHS}: Interpolated SED library, all normalised to the same value at 0.4$\mu$m; \textit{Center}: GOODS type-specific luminosity functions at $0.75<z<1$; \textit{RHS}: Fitted standard deviation of photometric redshift errors in each redshift slice of the simulation.
\label{fig:sim}}
\end{figure}

We take a model linear theory matter power spectrum and generate Gaussian realisations of over-densities.
Each over-density is assigned $N_i$ galaxies where $N_i$ depends on the value of the over-density 
and the expected mean number density of galaxies at the relevant redshift.  The mean number density of galaxies as a function of redshift is calculated by integrating the GOODS luminosity functions (LFs)\cite{goodslf}, shown in the center panel of Fig.~\ref{fig:sim}.

Each galaxy in the simulation is assigned an absolute magnitude and a broad spectral type: early-type, late-type or starburst.  This is done by creating a probability distribution using the galaxy-type specific LFs measured from GOODS\cite{goodslf}, again see the center panel of Fig.~\ref{fig:sim}.  We then create a library of spectral energy densities (SEDs) of different galaxy types by interpolating the CWW\cite{cww} and Kinney\cite{kinney} empirical galaxy templates, which cover the range of galaxy types from early to starburst, shown in the left-hand panel of Fig.~\ref{fig:sim}. We randomly assign each galaxy an SED with a prior based on its original broad spectral type.  Using the information simulated so far for each galaxy (specifically: $z$, absolute magnitude and SED) we calculate the observed apparent magnitudes in each LSST filter including the expected photometric and systematic errors\cite{lsstsb}.  We verify the simulation method by comparing our own simulation of the GOODS data to the real GOODS data.

Photometric redshifts are reconstructed using a $\chi^2$ fitting technique using the galaxy luminosity function as a prior. Our code produces reasonable distributions of the photometric redshift errors: $z_{phot}-z_{spec}$.  The right-hand panel of Fig.~\ref{fig:sim} shows the fitted standard deviation of $z_{phot}-z_{spec}$ in different redshift slices of $\Delta z=0.05$ in the simulation.  The precision after $z=1$ worsens  as expected due to the Balmer break having transitioned out of the $y$ band and the Lyman break having yet to enter the $u$ band.

\section{Early simulation result}
\label{sec:sim2}
We have simulated just a small survey at this stage: 5 Gpc$^3$ covering $0.7<z<1.4$ and assuming 100 visits of the simulated survey area.

We compute the power spectrum using a direct Fourier transform method as outlined in Blake et al. (2007)\cite{bmegaz}. To measure the precision on the BAO scale we use the ``wiggles only" \cite{bg03} method: dividing the measured power spectrum by a smooth reference power spectrum given by the Eisenstein and Hu\cite{eh98} ``no-wiggles" fitting formula.  The ``wiggles only" power spectrum is then well approximated by a decaying sinusoid:
\begin{equation}
\frac{P(k)}{P_{ref}} = 1 + A k\exp \left[ -\left(\frac{k}{0.1h\mbox{Mpc}^{-1}}\right)^\gamma\right]\sin\left(\frac{2\pi k}{k_a}\right)
\end{equation}
where $k_a$ represents the acoustic scale.
\begin{figure}
\center
\psfig{figure=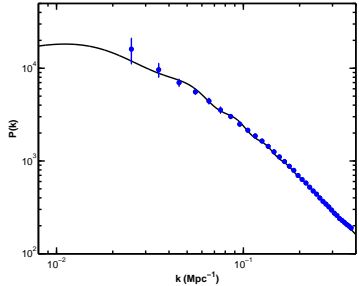,height=1.5in}
\caption{Power spectrum measured from the simulation described in Sections \ref{sec:sim1} and \ref{sec:sim2}.
\label{fig:ps}}
\end{figure}

We find the precision on the measurement of the acoustic scale from this simulation to be 3\%, corresponding to an error on the dark energy equation of state $w$ (assumed constant) of 16\%.  Performing a simple extrapolation to an equivalent simulation covering the LSST volume suggests the precision of the acoustic scale measurement for LSST from BAO would be 0.4\%, corresponding to an error on $w$ of about 2\%.

\section{LSST vs BigBOSS}

BigBOSS\cite{bigboss} is a proposed \textit{spectroscopic} survey to measure BAO.  Though it will have more precise redshifts than LSST it will cover less volume and observe less galaxies, see Table~\ref{tab:sur}. It is interesting to compare the expected errors on the power spectrum for idealised versions of both surveys.

An analytic expression for the expected fractional errors on the power spectrum under some simplifying assumptions is given below:
\begin{equation}
\label{eq:pserr}
\frac{\sigma_P}{P}=\sqrt{\frac{2(2\pi)^3}{V} \frac{1}{4\pi k^2 dk}}\left(\frac{P+1/n}{P}\right)\frac{1}{\left<\exp\left(-(k_z\sigma_z)^2\right>_k\right)}
\end{equation}
The above equation incorporates uncertainties due to sample variance, shot noise and the photometric redshift errors, assumed to have a Gaussian distribution with a standard deviation which is constant across the survey.  For BigBOSS essentially $\sigma_z=0$.

\begin{figure}
\center
\psfig{figure= 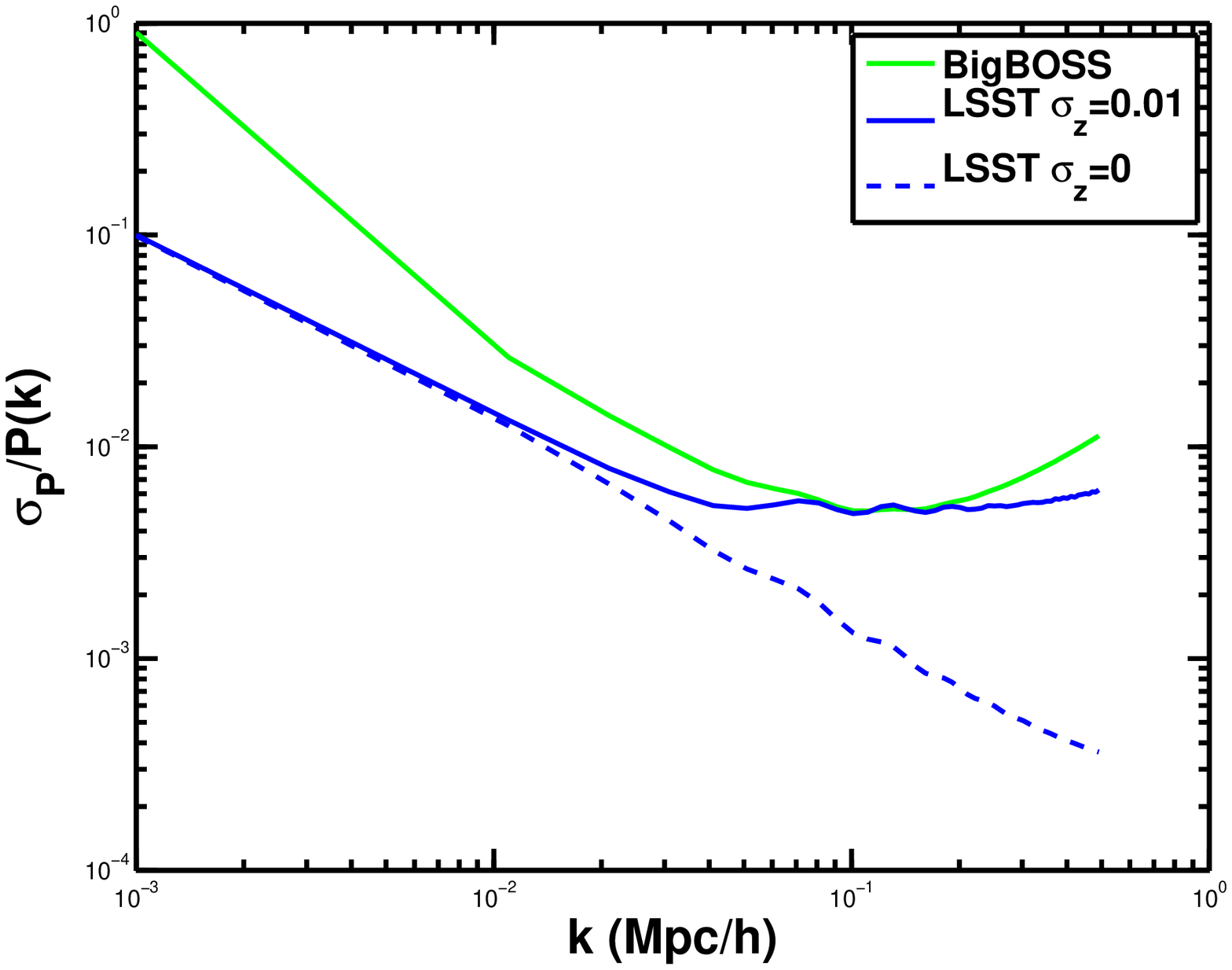,height=1.5in}
\psfig{figure= 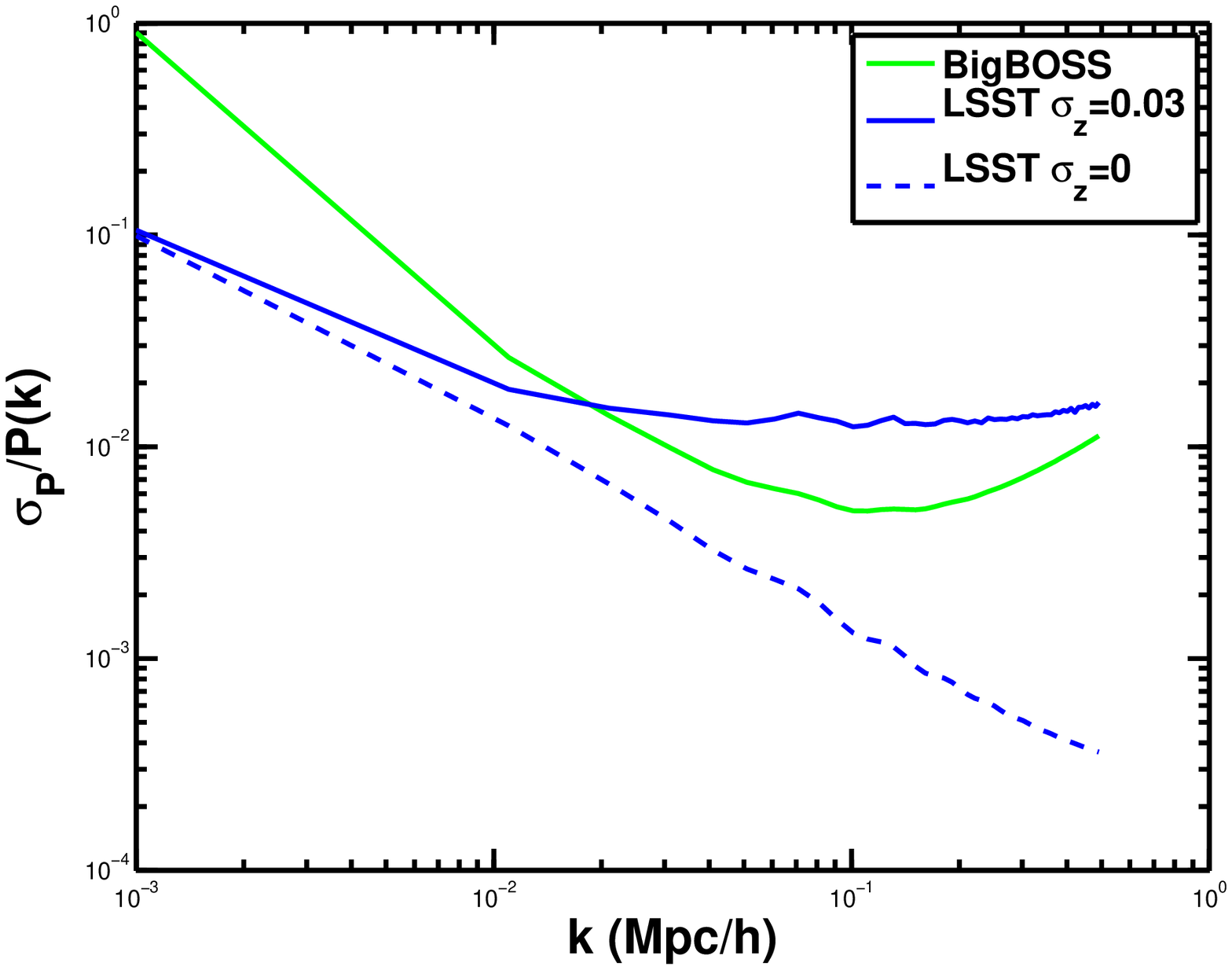,height=1.5in}
\caption{Expected power spectrum errors for BigBOSS and LSST as given by Eq.~\ref{eq:pserr}. On the left-hand side optimistic photometric redshift errors for LSST are assumed ($\sigma_z=0.01$), and on the right-hand side, more conservative ones ($\sigma_z=0.03$). 
\label{fig:pserr}}
\end{figure}

Fig.~\ref{fig:pserr} shows Eq.~\ref{eq:pserr} plotted for the LSST survey (blue/dark solid line) and the BigBOSS survey (green/light solid line).  On the left-hand side optimistic photometric redshift errors for LSST are assumed ($\sigma_z=0.01$), and on the right-hand side, conservative ones ($\sigma_z=0.03$).  The blue dashed line is for LSST with zero photometric redshift errors.  The figure shows that BigBOSS becomes shot noise dominated around the BAO scale. To produce competitive constraints with BigBOSS LSST must achieve a photometric redshift precision of $\sigma_z=0.01$, though LSST performs far superior to BigBOSS on large scales for both assumed error models.

\begin{table}
\caption{Comparison of BigBOSS and LSST\label{tab:sur}}
\begin{center}
\begin{tabular}{|c|c|c|c|c|}
\hline
Survey &  redshift range  & Sky Area (sq.deg) &Volume (Mpc/$h$)$^3$& N gals \\
\hline
BigBOSS & 0.20$<z<$2.0& 14,000 & 3x10$^{10}$ & 3x10$^7$ \\
\hline
LSST & 0.15$<z<$3.0& 20,000 & 1x10$^{11}$ & 10$^{10}$ \\
\hline
\end{tabular}
\end{center}
\end{table}

\section{Conclusions}

From our current simulations we find that:\\
\indent $\bullet$ Our simulation produces a realistic observed galaxy catalog.\\
\indent $\bullet$ Computation of photometric redshifts produces expected error distribution.\\
\indent $\bullet$ Using our current small survey simulation we have shown that a direct 3D  power spectrum\\ \indent \indent  estimation with photometric redshifts can produce a reasonable constraint on $k_{BAO}$: \indent \indent $\Delta k_a/k_a = 3$\% \\
\indent $\bullet$ LSST is competitive with BigBOSS even on the BAO scale, it is superior at large scales.\\
\indent \indent  $\Delta k_a/k_a = 0.3$\% (BigBOSS/LSST optimistic) \\
\indent \indent $\Delta k_a/k_a = 0.6$\% (LSST conservative) \\


\section*{References}
\begin{thebibliography}{99}
\bibitem{bb05}C. Blake, and S. Bridle \Journal{\MN}{363}{1329}{2005}.
\bibitem{bmegaz}C. Blake, A. Collister, S. Bridle and O. Lahav \Journal{\MN}{374}{1527}{2007}.
\bibitem{lsstsb} LSST Science Collaborations: {\em The LSST Science Book}, \Journal{\arxiv}{0912.0201v1}{http://www.lsst.org/lsst/scibook}{2009}.
\bibitem{goodslf}T. Dahlen et al. \Journal{\APJ}{631}{126}{2005}.
\bibitem{cww}G.D. Coleman, C.C. Wu and D.W. Weedman \Journal{\APJS}{43}{393}{1980}
\bibitem{kinney}A.L. Kinney et al. \Journal{\APJ}{467}{38}{1996}.
\bibitem{bg03}C. Blake, and K. Glazebrook \Journal{\APJ}{594}{665}{2003}.
\bibitem{eh98}D.J. Eisenstein and W. Hu \Journal{\APJ}{496}{605}{1998}.
\bibitem{bigboss}D.J. Schlegel et al. \Journal{\arxiv}{0904.0468}{}{2009}
\end{thebibliography}
\end{document}